\documentclass[apj]{emulateapj}
\usepackage{apjfonts}
\usepackage{amsmath}
\usepackage{appendix}

\shorttitle{Cosmology from ACT Sunyaev-Zel'dovich Galaxy Clusters}
\shortauthors{Sehgal et al.}

\begin{document}

\title{The Atacama Cosmology Telescope: Cosmology from Galaxy Clusters Detected via the Sunyaev-Zel'dovich Effect}

\author{
Neelima~Sehgal\altaffilmark{1},
Hy~Trac\altaffilmark{2,3},
Viviana~Acquaviva\altaffilmark{4,5},
Peter~A.~R.~Ade\altaffilmark{6},
Paula~Aguirre\altaffilmark{7},
Mandana~Amiri\altaffilmark{8},
John~W.~Appel\altaffilmark{9},
L.~Felipe~Barrientos\altaffilmark{7},
Elia~S.~Battistelli\altaffilmark{10,8},
J.~Richard~Bond\altaffilmark{11},
Ben~Brown\altaffilmark{12},
Bryce~Burger\altaffilmark{8},
Jay~Chervenak\altaffilmark{13},
Sudeep~Das\altaffilmark{14,9,4},
Mark~J.~Devlin\altaffilmark{15},
Simon~R.~Dicker\altaffilmark{15},
W.~Bertrand~Doriese\altaffilmark{16},
Joanna~Dunkley\altaffilmark{17,9,4},
Rolando~D\"{u}nner\altaffilmark{7},
Thomas~Essinger-Hileman\altaffilmark{9},
Ryan~P.~Fisher\altaffilmark{9},
Joseph~W.~Fowler\altaffilmark{9,16},
Amir~Hajian\altaffilmark{11,4,9},
Mark~Halpern\altaffilmark{8},
Matthew~Hasselfield\altaffilmark{8},
Carlos~Hern\'andez-Monteagudo\altaffilmark{18},
Gene~C.~Hilton\altaffilmark{16},
Matt~Hilton\altaffilmark{19,20},
Adam~D.~Hincks\altaffilmark{9},
Ren\'ee~Hlozek\altaffilmark{17},
David~Holtz\altaffilmark{9},
Kevin~M.~Huffenberger\altaffilmark{21},
David~H.~Hughes\altaffilmark{22},
John~P.~Hughes\altaffilmark{5},
Leopoldo~Infante\altaffilmark{7},
Kent~D.~Irwin\altaffilmark{16},
Andrew~Jones\altaffilmark{9},
Jean~Baptiste~Juin\altaffilmark{7},
Jeff~Klein\altaffilmark{15},
Arthur~Kosowsky\altaffilmark{12},
Judy~M.~Lau\altaffilmark{1,23,9},
Michele~Limon\altaffilmark{24,15,9},
Yen-Ting~Lin\altaffilmark{25,4,7},
Robert~H.~Lupton\altaffilmark{4},
Tobias~A.~Marriage\altaffilmark{4,26},
Danica~Marsden\altaffilmark{15},
Krista~Martocci\altaffilmark{27,9},
Phil~Mauskopf\altaffilmark{6},
Felipe~Menanteau\altaffilmark{5},
Kavilan~Moodley\altaffilmark{19,20},
Harvey~Moseley\altaffilmark{13},
Calvin~B.~Netterfield\altaffilmark{28},
Michael~D.~Niemack\altaffilmark{16,9},
Michael~R.~Nolta\altaffilmark{11},
Lyman~A.~Page\altaffilmark{9},
Lucas~Parker\altaffilmark{9},
Bruce~Partridge\altaffilmark{29},
Beth~Reid\altaffilmark{9,30},
Blake~D.~Sherwin\altaffilmark{9},
Jon~Sievers\altaffilmark{11},
David~N.~Spergel\altaffilmark{4},
Suzanne~T.~Staggs\altaffilmark{9},
Daniel~S.~Swetz\altaffilmark{15,16},
Eric~R.~Switzer\altaffilmark{27,9},
Robert~Thornton\altaffilmark{15,31},
Carole~Tucker\altaffilmark{6},
Ryan~Warne\altaffilmark{19},
Ed~Wollack\altaffilmark{13},
Yue~Zhao\altaffilmark{9}
}
\altaffiltext{1}{Kavli Institute for Particle Astrophysics and Cosmology, Stanford
University, Stanford, CA, USA 94305-4085}
\altaffiltext{2}{Department of Physics, Carnegie Mellon University, Pittsburgh, PA 15213}
\altaffiltext{3}{Harvard-Smithsonian Center for Astrophysics, 
Harvard University, Cambridge, MA, USA 02138}
\altaffiltext{4}{Department of Astrophysical Sciences, Peyton Hall, 
Princeton University, Princeton, NJ USA 08544}
\altaffiltext{5}{Department of Physics and Astronomy, Rutgers, 
The State University of New Jersey, Piscataway, NJ USA 08854-8019}
\altaffiltext{6}{School of Physics and Astronomy, Cardiff University, The Parade, Cardiff, Wales, UK CF24 3AA}
\altaffiltext{7}{Departamento de Astronom{\'{i}}a y Astrof{\'{i}}sica, 
Facultad de F{\'{i}}sica, Pontific\'{i}a Universidad Cat\'{o}lica de Chile,
Casilla 306, Santiago 22, Chile}
\altaffiltext{8}{Department of Physics and Astronomy, University of
British Columbia, Vancouver, BC, Canada V6T 1Z4}
\altaffiltext{9}{Joseph Henry Laboratories of Physics, Jadwin Hall,
Princeton University, Princeton, NJ, USA 08544}
\altaffiltext{10}{Department of Physics, University of Rome ``La Sapienza'', 
Piazzale Aldo Moro 5, I-00185 Rome, Italy}
\altaffiltext{11}{Canadian Institute for Theoretical Astrophysics, University of
Toronto, Toronto, ON, Canada M5S 3H8}
\altaffiltext{12}{Department of Physics and Astronomy, University of Pittsburgh, 
Pittsburgh, PA, USA 15260}
\altaffiltext{13}{Code 553/665, NASA/Goddard Space Flight Center,
Greenbelt, MD, USA 20771}
\altaffiltext{14}{Berkeley Center for Cosmological Physics, LBL and
Department of Physics, University of California, Berkeley, CA, USA 94720}
\altaffiltext{15}{Department of Physics and Astronomy, University of
Pennsylvania, 209 South 33rd Street, Philadelphia, PA, USA 19104}
\altaffiltext{16}{NIST Quantum Devices Group, 325
Broadway Mailcode 817.03, Boulder, CO, USA 80305}
\altaffiltext{17}{Department of Astrophysics, Oxford University, Oxford, 
UK OX1 3RH}
\altaffiltext{18}{Max Planck Institut f\"ur Astrophysik, Postfach 1317, 
D-85741 Garching bei M\"unchen, Germany}
\altaffiltext{19}{Astrophysics and Cosmology Research Unit, School of
Mathematical Sciences, University of KwaZulu-Natal, Durban, 4041,
South Africa}
\altaffiltext{20}{Centre for High Performance Computing, CSIR Campus, 15 Lower
Hope St. Rosebank, Cape Town, South Africa}
\altaffiltext{21}{Department of Physics, University of Miami, Coral Gables, 
FL, USA 33124}
\altaffiltext{22}{Instituto Nacional de Astrof\'isica, \'Optica y 
Electr\'onica (INAOE), Tonantzintla, Puebla, Mexico}
\altaffiltext{23}{Department of Physics, Stanford University, Stanford, CA, 
USA 94305-4085}
\altaffiltext{24}{Columbia Astrophysics Laboratory, 550 W. 120th St. Mail Code 5247,
New York, NY USA 10027}
\altaffiltext{25}{Institute for the Physics and Mathematics of the Universe, 
The University of Tokyo, Kashiwa, Chiba 277-8568, Japan}
\altaffiltext{26}{Dept. of Physics and Astronomy, The Johns Hopkins University, 3400 N. Charles St., Baltimore, MD 21218-2686}
\altaffiltext{27}{Kavli Institute for Cosmological Physics, 
Laboratory for Astrophysics and Space Research, 5620 South Ellis Ave.,
Chicago, IL, USA 60637}
\altaffiltext{28}{Department of Physics, University of Toronto, 
60 St. George Street, Toronto, ON, Canada M5S 1A7}
\altaffiltext{29}{Department of Physics and Astronomy, Haverford College,
Haverford, PA, USA 19041}
\altaffiltext{30}{Institut de Ciencies del Cosmos (ICC), University of
Barcelona, Barcelona 08028, Spain}
\altaffiltext{31}{Department of Physics , West Chester University 
of Pennsylvania, West Chester, PA, USA 19383}

\newcommand{\signomarg}{0.821 \pm 0.044}
\newcommand{\sigwmarg}{0.851\pm 0.115}
\newcommand{\wnomarg}{-1.05 \pm  0.20}
\newcommand{\wwmarg}{-1.14 \pm 0.35}
\newcommand{\Aparam}{8.77 \pm 3.77}
\newcommand{\Bparam}{1.75 \pm 0.28}
\newcommand{\Cparam}{0.97 \pm 0.68}
\newcommand{\Sparam}{0.27 \pm 0.13}

\newcommand{\arone}{148\,GHz}
\newcommand{\area}{455 square degrees}

\def\S{Section }
\def\mk{ \rm{\mu K}}
\newcommand{\totalClusters}{9}
\newcommand{\specClusters}{Six}
\newcommand{\photoClusters}{three}
\newcommand{\ymin}{300 $\mk$}
\newcommand{\yerror}{60 $\mk$}
\newcommand{\snr}{5}

\newcommand{\sn}{y}
\newcommand{\snobs}{y^{\rm{obs}}}
\newcommand{\sntrue}{y^{\rm{true}}}
\newcommand{\m}{M^{\rm{true}}}
\newcommand{\zobs}{z^{\rm{obs}}}
\newcommand{\ztrue}{z^{\rm{true}}}
\newcommand{\lnm}{{\rm ln}M^{\rm{true}}}
\newcommand{\pr}{P_r}
\newcommand{\cosm}{\{ c_j \}}
\newcommand{\s}{S}
\newcommand{\Msunh}{M_{\odot} h^{-1}}
\newcommand{\n}{\noindent}
\newcommand{\lcdm}{\Lambda \rm{CDM}}
\def\Msun{M_\odot}
\def\sig{\sigma_8}
\def\ltsima{$\; \buildrel < \over \sim \;$}
\def\simlt{\lower.5ex\hbox{\ltsima}}
\def\gtsima{$\; \buildrel > \over \sim \;$}
\def\simgt{\lower.5ex\hbox{\gtsima}}
\def\simless{\mathbin{\lower 3pt\hbox
   {$\rlap{\raise 5pt\hbox{$\char'074$}}\mathchar"7218$}}}   % < or of order
\def\simgreat{\mathbin{\lower 3pt\hbox
   {$\rlap{\raise 5pt\hbox{$\char'076$}}\mathchar"7218$}}}   % > or of order
\def\Mpch{h^{-1}{\rm Mpc}}

\begin{abstract}
We present constraints on cosmological parameters based on a sample of Sunyaev-Zel'dovich-selected galaxy clusters detected in a millimeter-wave survey by the Atacama Cosmology Telescope.  The cluster sample used in this analysis consists of \totalClusters\ optically-confirmed high-mass clusters comprising the high-significance end  of the total cluster sample identified in \area\ of sky surveyed during 2008 at 148\,GHz.  We focus on the most massive systems to reduce the degeneracy between unknown cluster astrophysics and cosmology
derived from SZ surveys.  We describe the scaling relation between cluster mass and SZ signal with a 4-parameter fit.  Marginalizing over the values of the parameters in this fit with conservative priors gives $\sigma_8=\sigwmarg$ and $w=\wwmarg$ for a spatially-flat wCDM cosmological model with WMAP 7-year priors on cosmological parameters.  This gives a modest improvement in statistical uncertainty over WMAP 7-year constraints alone.  Fixing the scaling relation between cluster mass and SZ signal to a fiducial relation obtained from numerical simulations and calibrated by X-ray observations, we find $\sigma_8=  \signomarg$ and $w=\wnomarg$.  These results are consistent with constraints from WMAP 7 plus baryon acoustic oscillations plus type Ia supernoava which give $\sigma_8 = 0.802\pm0.038$ and $w= -0.98 \pm 0.053$.  A stacking analysis of the clusters in this sample compared to clusters simulated assuming the fiducial model also shows good agreement.  These results suggest that, given the sample of clusters used here, both the astrophysics of massive clusters and the cosmological parameters derived from them are broadly consistent with current models.
\end{abstract}


\section{INTRODUCTION} \label{sec:intro}

Ever-improving observations suggest a concordant picture of our Universe.  In this picture, generally called $\lcdm$, "dark energy," the component responsible for the Universe's  accelerated expansion, is believed to be the energy of the vacuum with a constant equation of state parameter, $w$, equal to $-1$ \citep[e.g.,][]{Riess2009,Brown2009,Hicken2009,Kessler2009,Percival2010,Komatsu2010}.  $\lcdm$ has been measured via probes of the Universe's expansion rate such as type Ia supernovae, the primary cosmic microwave background, and baryon acoustic oscillations.  However, $\lcdm$ also makes concrete predictions about the Universe's growth of structure.  This growth rate describes how quickly dark matter halos form and evolve over cosmic time.  A deviation from this predicted growth rate, particularly on linear scales, would signal a breakdown of $\lcdm$ (see e.g., \citet{Linder2005, Bertschinger2008, Silvestri2009, Jain2010, Shapiro2010} and references therein).  

A handful of techniques have been available for measuring the growth of structure in the Universe.  These largely consist of observing the weak and strong lensing of background sources by intervening matter \citep[e.g.,][]{Schrabback2010}, measuring distortions in redshift space with spectroscopic surveys of galaxies \citep[e.g.,][]{Simpson2010}, and quantifying the abundance of galaxy clusters as a function of mass and redshift \citep[e.g.,][]{Bahcall1998}).  This latter technique is one of the oldest and has been maturing with the advent of large area X-ray  \citep[e.g.,][]{Truemper1990} and optical  \citep[e.g.,][]{Koester2007} surveys.

Millimeter-wave surveys now possess the resolution and sensitivity to detect galaxy clusters.  Detecting galaxy clusters via the Sunyaev-Zel'dovich (SZ) effect in large area millimeter-wave maps, as have become available through the Atacama Cosmology Telescope (ACT) \citep{Swetz2010} and the South Pole Telescope (SPT) \citep{Carlstrom2009}, is a potentially powerful method.  Cluster selection using the SZ effect \citep{Zeldovich1969,SZ1970, SZ1972} is the technique whose selection function is least dependent on cluster redshift.  This allows for a complete picture of the evolution of clusters from their first formation to the present.

Here we probe structure growth with a measurement of the abundance of massive galaxy clusters from observations made by the ACT project in 2008.  We focus on the most massive SZ-selected clusters as this is the regime where high signal-to-noise measurements exist and we can best understand the cluster astrophysics.  We also note that the clusters considered in this work are rare and represent the tail of the mass distribution, which is sensitive to the background cosmology.  This analysis uses the number of massive galaxy clusters to constrain, in particular, the normalization of the matter power spectrum, $\sig$, and the dark energy equation-of state-parameter, $w$. 

This paper is structured as follows: \S \ref{sec:background} describes the SZ effect and the ACT SZ cluster survey.  \S \ref{sec:catalog} describes the 2008 ACT high-significance cluster sample. In \S \ref{sec:results}, we present our results, and in \S \ref{sec:discuss}, we discuss their implications and conclude.

\section{Background} \label{sec:background}

\subsection{The Thermal Sunyaev-Zel'dovich Effect} \label{sec:sz}

The thermal SZ effect arises when primary cosmic microwave background photons, on their path from the last scattering surface, encounter an intervening galaxy cluster.  The hot ionized gas within the cluster inverse Compton scatters about $1\%$ of the CMB photons, boosting their energy and altering the intensity of the microwave background as a function of frequency at the location of the cluster \citep{SZ1970, SZ1972}.   To first order, the effective temperature shift (which is proportional to the intensity shift), at a frequency $\nu$, from the thermal SZ effect is given by 

\begin{equation}
\frac{\Delta T}{T_{\rm CMB}} =  f(x) \frac{k_B \sigma_T}{m_e c^2}\int n_e T_e {\rm d}l \equiv f(x) y,
\label{eqn:tsz}
\end{equation}
where $n_e$ and $T_e$ are the number density and temperature of the electron distribution of the cluster gas, ${\rm d}l$ is the line-of-sight path length through the cluster, $ \sigma_T$ is the Thomson cross-section,  $k_B$ is the Boltzmann constant, and 

\begin{equation}
f(x) \equiv x\coth(x/2) - 4 , \quad x\equiv h\nu/(k_{\rm B}T_{\rm CMB}) .
\end{equation}
Here $y$ is the usual Compton y-parameter.  Note that the full SZ effect contains relativistic corrections as in \citet{Nozawa1998}, which we include in our simulations.  For our sample, these corrections are 5 to 10\%.  We take Eq.~\ref{eqn:tsz}, describing the first-order thermal SZ effect, as the definition of $y$ we use throughout this work, and treat the relativistic corrections as an additional source of noise (see \S \ref{subsec:Yrecov}).

At frequencies below 218 GHz, where the signal is null, $\Delta T$ is negative, and the cluster appears as a cold spot in CMB maps.  Above the null, $\Delta T$ is positive, and the cluster appears as a hot spot.  Eq.~\ref{eqn:tsz} is also redshift independent, and the amplitude of the intensity shift is to first order proportional only to the thermal pressure of the cluster.  This makes the SZ effect especially powerful for two reasons: the microwave background can trace all the clusters of a given thermal pressure that have formed between the last scattering surface and today in a redshift independent way, and the amplitude of this effect, being proportional to the thermal pressure, is closely related to the cluster mass. 

\subsection{The ACT Sunyaev-Zel'dovich Cluster Survey}\label{sec:survey}

The Atacama Cosmology Telescope (ACT) is a 6-meter off-axis telescope designed for arcminute-scale millimeter-wave observations \citep{Swetz2010, Hincks2009}.  It is located on Cerro Toco in the Atacama Desert of Chile.  One goal of this instrument is to measure the evolution of structure in the Universe via the SZ effect.  In the 2008 observing season ACT surveyed \area\ of sky in the southern hemisphere at 148\,GHz.  In this survey, galaxy clusters were detected from their SZ signal (see \citet{Marriage2010clusters} for details).   A sample of 23 SZ-selected clusters was optically confirmed using multi-band optical imaging on 4-meter telescopes during the 2009B observing season (see \citet{Menanteau2010b} for details). Some of the low-redshift systems in this sample are previously known clusters for which spectroscopic redshifts are available. However, roughly half are newly detected systems, and photometric redshift estimates have been obtained from optical imaging.  Here we make use of the subsample of these clusters with high-significance SZ detections (signal-to-noise ratio $>\snr$, as defined in \S \ref{subsec:detectMeth}) to obtain cosmological parameter constraints.  

\section{THE 2008 ACT HIGH-SIGNIFICANCE CLUSTER CATALOG} \label{sec:catalog}

\subsection{CMB Data} \label{sec:data}

Here we give a brief overview of the survey observations and the reduction of the raw data to maps.  For a more complete introduction to the ACT instrument, observations, and data reduction pipeline, we refer the reader to \citet{Fowler2010} and \citet{Swetz2010}.   The 2008 observations in the southern hemisphere were carried out between mid-August and late December over a 9$^\circ$ wide ACT strip centered on a declination of $-$53$^\circ$ degrees and extending from approximately 19$^h$ through 0$^h$ to 7$^h$30 in right ascension. The \area\ used for this analysis consists of a 7$^\circ$-wide strip centered at a declination of $-$52$^\circ 30^\prime$ and running from right ascension $00^h 12^m$ to $7^h 10^m$. The resolution of the ACT instrument is about $1.4 \arcmin$ at 148\,GHz.  Typical noise levels in the map are 30 $\rm{\mu K}$ per square arcminute, rising to 50 $\rm{\mu K}$ toward the map boundaries.  Seven of the nine clusters considered in this work fall in the central region of the map with lower noise levels.

Rising and setting scans cross-link each point on the sky with adjacent points such that the data contain the information necessary to make a map recovering brightness
fluctuations over a wide range of angular scales.  In addition to survey observations, ACT also executed regular observations of Uranus and Saturn during 2008 to provide beam profiles, pointing, and temperature calibration.  Analysis of the beam profiles is discussed in \citet{Hincks2009}.  Absolute pointing is determined by comparing the positions of ACT-observed radio sources with the positions of these same sources detected in the AT20G survey \citep{Murphy2010}.  The final temperature calibration at 148\,GHz is based on a recent analysis cross-correlating ACT and WMAP maps and is determined with an uncertainty of 2\% (Hajian et al. 2010). The small residual calibration uncertainty translates into a small systematic uncertainty in $y$ values of observed clusters, which is negligible compared to other uncertainties discussed in this analysis. 

To make maps, an iterative preconditioned conjugate gradient solver is used to recover the maximum likelihood maps. This algorithm solves simultaneously for the millimeter sky as well as correlated noise (e.g., a common mode from atmospheric emission).  The map projection used is cylindrical equal area with a standard latitude of $-$53$^\circ 30^\prime$ and square pixels, $0.5\arcmin$ on a side. 

\begin{center}
\begin{deluxetable*}{cccccc}
\tabletypesize{\scriptsize}
\tablecaption{ACT Cluster Catalog for High-Significance Clusters from the 2008 Observing Season\label{tab:clustProperties}}
\tablehead{\colhead{ACT Descriptor} & \colhead{R.A.} & \colhead{decl.} & \colhead{$\sn T_{\rm{CMB}} (\rm{\mu K})$\tablenotemark{$\dagger$}} & \colhead{Redshift} & \colhead{Other Name}}
\startdata
 ACT-CL~J0645$-$5413 & 06:45:30 & $-$54:13:39 & 340 $\pm$ 60&  0.167\tablenotemark{a} & Abell 3404\\
 ACT-CL~J0638$-$5358 & 06:38:46  & $-$53:58:45 & 540 $\pm$ 60&  0.222\tablenotemark{a} & Abell S0592\\ 
 ACT-CL~J0658$-$5557 & 06:58:30 & $-$55:57:04 & 560 $\pm$ 60&  0.296\tablenotemark{b} & 1ES0657$-$558(Bullet)\\ 
 ACT-CL~J0245$-$5302 & 02:45:33  & $-$53:02:04 & 475 $\pm$ 60&  0.300\tablenotemark{c} &  Abell S0295\\
 ACT-CL~J0330$-$5227 & 03:30:54  & $-$52:28:04 & 380 $\pm$ 60&  0.440\tablenotemark{d} & Abell 3128(NE)\\
 ACT-CL~J0438$-$5419 & 04:38:19  & $-$54:19:05 & 420 $\pm$ 60&  $0.54\pm0.05$\tablenotemark{e}& New\\
 ACT-CL~J0616$-$5227 & 06:16:36  & $-$52:27:35 & 360 $\pm$ 60&  $0.71\pm0.10$\tablenotemark{e}& New\\
 ACT-CL~J0102$-$4915 & 01:02:53  & $-$49:15:19 & 490 $\pm$ 60&  $0.75\pm0.04$\tablenotemark{e}& New\\
 ACT-CL~J0546$-$5345 & 05:46:37  & $-$53:45:32 & 310 $\pm$ 60&  1.066\tablenotemark{f} &  ~~~~SPT-CL 0547$-$5345 
\enddata
\tablenotetext{$\dagger$}{$\rm{\mu K}$ given for the brightest $0.5\arcmin$ pixel of each cluster}
\tablenotetext{a}{spectroscopic-z from \cite{deGrandi1999}}
\tablenotetext{b}{spectroscopic-z from \cite{Tucker1998}}
\tablenotetext{c}{spectroscopic-z from \cite{Edge1994}}
\tablenotetext{d}{spectroscopic-z from \cite{Werner2007}}
\tablenotetext{e}{photometric-z from \cite{Menanteau2010b}}
\tablenotetext{f}{spectroscopic-z from \cite{Infante2010,Brodwin2010}}
\end{deluxetable*}
\end{center}

\vspace{-1.0cm}
\subsection{Cluster Detection Method}\label{subsec:detectMeth}

In order to detect clusters in single-frequency millimeter-wave maps we construct a filter that is similar in morphology to the clusters we are trying to detect.  We adopt a matched filter of the form

\begin{equation}
\psi({\bf{k}}) = \bigg[ \frac{1}{(2\pi)^2}\int \frac{|\tau({\bf{k'}})|^2}{P({\bf{k'}})} d^2k'  \bigg]^{-1} \frac{\tau({\bf{k}})}{P({\bf{k}})}
\label{eqn:filter}
\end{equation}
following \citet{Haehnelt1996}, \citet{Herranz2002a,Herranz2002b}, and \citet{Melin2006}.  Here $\tau({\bf{k}})$ is the beam convolved cluster signal in Fourier space, and $P({\bf{k}})$ is the power spectrum of the noise, both astrophysical and instrumental.  The astrophysical noise sources for cluster detection include the primary CMB lensed by intervening structure, radio galaxies, dusty star-forming galaxies, Galactic dust, and the SZ background from unresolved clusters, groups, and the intergalactic medium.  Since the power from the SZ signal is subdominant to these astrophysical sources (as evidenced by \citet{Lueker2009}, \citet{Hall2010}, \citet{Fowler2010}, \citet{Das2010}, and \citet{Dunkley2010}), we can to a good approximation model the power spectrum of the total noise as the power spectrum of the data itself.  In Eq.~\ref{eqn:filter}, the quantity in square brackets serves as a normalization factor to ensure an unbiased estimate of the cluster signal.  When multi-frequency maps are available, this filter can be modified to incorporate the known spectral signature of the SZ signal.

The template shape that we choose to match the cluster morphology is given by a two-dimensional Gaussian profile, which in Fourier space has the form\setcounter{footnote}{0}\footnote{We use the flat space approximation, ${\bf{l}} = 2 \pi {\bf{k}}$.}  
 
\begin{equation}
\Delta T({l}) = A{\rm{exp}} [-\theta^2 (l+1)l/2].  
\label{eqn:filter2}
\end{equation}
Here $\theta = {\rm{FWHM}}/ \sqrt{8 {\rm{ln}} 2}$ where FWHM is the full width at half maximum, and A is a normalization factor that will be derived from
simulations (see \S \ref{subsec:Yrecov}).  We choose FWHM to be $2 \arcmin$ as this is a typical cluster size in our maps.  The analysis presented here of the cosmological parameters is nearly independent of the particular profile chosen for the cluster template, as long as the template is smooth and well-matched to the cluster angular size.

Before filtering our map to find clusters, we multiply the map, pixel-wise, by the square root of the number of observations per pixel normalized by the observations per pixel in the deepest part of the map in order to establish uniform noise properties.  We then detect point sources (radio and infrared galaxies) by a matched filter with the ACT beam as the template.  Selecting all point sources with a signal-to-noise ratio greater than $4.0$ in this filtered map, we mask them by replacing all on-source pixels with signal-to-noise ratio greater than $4.0$ with the average of the brightness in an annulus $4\arcmin$ away from the source center.  We do this  to avoid false detections due to the filter ringing around bright sources.  See \citet{Marriage2010sources} for details regarding point source detection. 

After masking out the brightest point sources, we filter the map to find clusters.  Clusters are then detected within this filtered map with a simple peak detection algorithm along the lines of SExtractor \citep{Bertin1996}.  An SZ $y$ value for the brightest $0.5 \arcmin$ pixel is measured for each cluster using this filtered map.  This definition of $y$ is different from the {\it{integrated}} $Y$, which is specifically the Compton-y parameter integrated over the face of the cluster and given by $Y=\int y d\Omega$.  The integration for this $Y$ value is performed over a radius tied to the size of the cluster, and a $Y$ defined this way would be a preferable quantity to use, having lower scatter with mass in theory \citep[e.g.,][]{daSilva2004,Motl2005,Nagai2006,Reid2006,Bonaldi2007}.  However, given single-frequency millimeter-wave maps, the size of each cluster cannot always be robustly determined.  An alternative quantity to measure is a  "central $y$ value," generally referred to as $y_0$, which essentially describes the normalization of the specific template shape used to find the cluster.  This quantity is not ideal for a cosmological analysis as it is intimately tied to the profile shape whereby the $y_0$ value is determined.  These values also exhibit a larger scatter with cluster mass than an integrated $y$ quantity \citep[e.g.,][]{Motl2005}. For all the clusters considered in this analysis, we fix an aperture size, given by our pixel size of $0.5\arcmin$, and measure Compton-y values within this fixed aperture. We do not consider larger aperture sizes here because we wish to limit template shape and redshift dependence.

We find that selecting clusters with a $yT_{\rm{CMB}}$ value\footnote{Note that $y$ is a dimensionless parameter.  We multiply it by $T_{\rm{CMB}} = 2.726 \times 10^6~\mk$ to give an indication of the expected temperature decrements.  For the frequency dependence, $f(x) \approx -1$ in Eq.~\ref{eqn:tsz} at 148\,GHz.} greater than \ymin\ corresponds to a subsample of clusters with a signal-to-noise ratio greater than $\snr$.  Here signal-to-noise ratio is defined as the signal of the brightest cluster pixel in the filtered map divided by the square root of the noise variance in the filtered map.  This subsample corresponds to the subsample of clusters with signal-to-noise ratio $\geq$ 5.9 in \citet{Marriage2010clusters}.\footnote{In  \citet{Marriage2010clusters}, a different detection method is used that varies the angular scale of the filter to match clusters of different sizes, and assigns a signal-to-noise ratio based on the scale that gives the highest value. The one cluster that has a signal-to-noise ratio $\geq$ 5.9 in that work that is not included here is ACT-CL J0235-5121.  In \citet{Marriage2010clusters}, it was found to have a high signal-to-noise ratio using a template scale of $4.0\arcmin$.  Although there is no doubt that this is a massive cluster (Menanteau et al. 2010a), its relatively high redshift, $z = 0.43 \pm 0.07$ argues for a compact size, suggesting that CMB contamination could be boosting the clusters's signal-to-noise ratio on a $4.0^\prime$ scale, as discussed in \citet{Marriage2010clusters}. This cluster is not found with signal-to-noise ratio $> 5$ using the $2\arcmin$ FWHM Gaussian template described above.}  The $y$ values derived with the
method used in this paper are not directly comparable to those in \citet{Marriage2010clusters}, in which an optimal filter sized is searched for. However, the methods are independently compared to simulations. While we detect clusters down to a signal-to-noise ratio of about 3 as defined in \citet{Marriage2010clusters}, we use only this higher-significance subsample in this work.  This subsample is given in Table \ref{tab:clustProperties}.

\subsection{Simulations and SZ Signal Recovery}\label{subsec:Yrecov}

To determine the expected scatter in our recovered $yT_{\rm{CMB}}$ values, we perform the same detection procedure discussed above on simulated maps.  Hereafter, the simulations we refer to are those discussed in \citet{Sehgal2010}, which include the SZ signal, lensed primary cosmic microwave background, Galactic dust, and radio and infrared sources correlated with SZ clusters as suggested by observations.  The large-scale structure in this simulation was carried out using a tree-particle-mesh code \citep{bode.ostriker.ea:2000, bode.ostriker:2003}, with a simulation volume of $1000\ \Mpch$ on a side containing $1024^{3}$ particles.   The cosmology adopted is consistent with the WMAP 5-year results \citep{Komatsu2009} though the details of the cluster properties are relatively insensitive to the background cosmology.  The mass distribution covering one octant of the full sky was saved, and halos with a friends-of-friends mass above $1 \times 10^{13} M_{\odot}$ and with a redshift below $z=3$ are identified.  The thermal SZ signal is derived by adding to the N-body halos a gas prescription that assumes a polytropic equation of state and hydrostatic equilibrium.  This model, which is described in more detail in \citet{Bode2009}, adjusts four free parameters (star-formation rate, nonthermal pressure support, dynamical energy transfer, and feedback from active galactic nuclei) which are calibrated against X-ray gas fractions as a function of temperature from the sample of \citet{Sun2009} and \citet{Vikhlinin06}.   The pressure profiles of the massive, low-redshift clusters in this simulation agree well with the best-fit profile of \citet{Arnaud2009} based on X-ray observations of high-mass, low-redshift systems \citep{Trac2010}.   We will see in \S \ref{sec:stacking} that the stacked SZ signal of the clusters in Table \ref{tab:clustProperties} is also consistent with the stacked thermal SZ signal of the massive clusters in this simulation.  The kinetic SZ in this simulation is calculated from the line-of-sight momentum of the particles.  We also include the relativistic corrections to the SZ signal as given in \citet{Nozawa1998}.  We convolve these simulations with the ACT beam and run them through the same map-making process discussed in \S \ref{sec:data}, including simulated atmospheric emission and realistic instrumental noise.  

\begin{figure}[t]
\epsscale{1.1}
\plotone{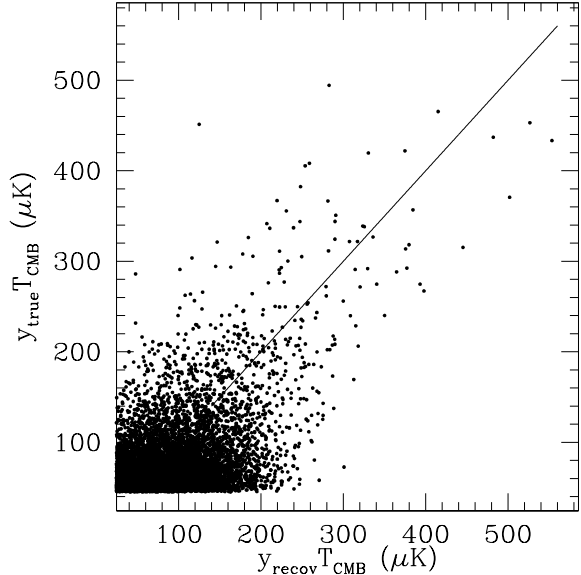}
\caption{\label{fig:Yrecov} The true $y$ value (as defined in Eq.~\ref{eqn:tsz}) for the brightest $0.5 \arcmin$ pixel of each cluster versus the recovered $y$ value from simulations using the detection method outlined in \S \ref{subsec:detectMeth}.  The dimensionless $y$ values have been multiplied by $T_{\rm{CMB}} = 2.726 \times 10^6~\mk$ to give an indication of the expected temperature decrements at 148\,GHz.  The root-mean-square scatter shown here is \yerror.}
\end{figure}

From these simulations, we cut out six different patches of \area\ to mimic the sky coverage in this analysis.  These six sky patches give us about 40 clusters that would correspond to the high-significance cluster sample given in Table \ref{tab:clustProperties}.  Using these simulations, we apply the same cluster detection procedure as discussed in \S \ref{subsec:detectMeth}, and recover $yT_{\rm{CMB}}$ values for the detected clusters. These recovered $yT_{\rm{CMB}}$ values are compared to the true $yT_{\rm{CMB}}$ values taken from the first-order thermal SZ maps alone, prior to any instrumental or atmospheric modifications.  The comparison between these true and recovered $yT_{\rm{CMB}}$ values is shown in Figure \ref{fig:Yrecov}.  We set the normalization factor, $A$, in Eq.~\ref{eqn:filter2} such that the mean bias between true and recovered $yT_{\rm{CMB}}$ values is zero.  The root-mean-square scatter in the recovered $yT_{\rm{CMB}}$ values is 60 $\mk$.  The scatter here is dominated by the instrumental and atmospheric noise sources in the map and not by the technique itself.  This same normalization is used for filtering the data map, and the same scatter is assumed.  Figure \ref{fig:Yrecov} also shows that any bias due to boosting  a cluster with an intrinsic $yT_{\rm{CMB}}<$ \ymin\ to a value above \ymin\ is far below the scatter. \citet{Vanderlinde2010} found that this signal boosting effect is at most 4\% for clusters detected with a signal-to-noise ratio greater than 5. 

\subsection{Optical Identification and Cluster Redshifts}\label{sec:redshifts}

The sample of SZ-selected clusters obtained from the data via the method above was followed up with optical observations to verify the millimeter-wave cluster identifications and determine cluster redshifts.  Here we provide a summary of the observing strategy and redshift determinations of our cluster sample, and we refer the
reader to \cite{Menanteau2010b} for a detailed description.

SZ cluster candidates were observed during the 2009B observing season with optical imaging on the 4-meter SOAR and NTT telescopes to search for a brightest
cluster galaxy and an accompanying red sequence of cluster
members.  While some of the clusters in the SZ-selected sample correspond to
previously known systems at low redshift ($z\simless0.3$), some
represent new systems, previously undetected at other wavelengths. 

Photometric redshifts and their probability distributions, $p(z)$,
were computed for each object from their dust-corrected $gri$
isophotal magnitude using the BPZ code \citep{Benitez2000}. \specClusters\ of the
clusters in our sample have spectroscopic redshift information
available (see Table \ref{tab:clustProperties} for references), as they were previously
known systems.
For the remaining \photoClusters\ systems we provide photometric
redshifts based on the NTT and SOAR imaging. Table \ref{tab:clustProperties} gives the mean
photometric redshift for these clusters obtained by iteratively selecting galaxies photometrically
classified as E or E/S0s within a projected radius of $500$~kpc and
redshift interval $|\Delta z|=0.05$.  This was done to obtain a local
color-magnitude relation for each color combination using a $3\sigma$
median sigma-clipping algorithm.  The uncertainties on the photometric redshifts of the three new clusters
in Table \ref{tab:clustProperties} come from the weighted rms of the individual galaxies
chosen as members.  Details of the photo-z algorithm are given in \cite{Menanteau2010a}.

\section{RESULTS}\label{sec:results}

\subsection{SZ Selection Function}\label{sec:selection}

\begin{figure}[t]
\epsscale{1.1}
\plotone{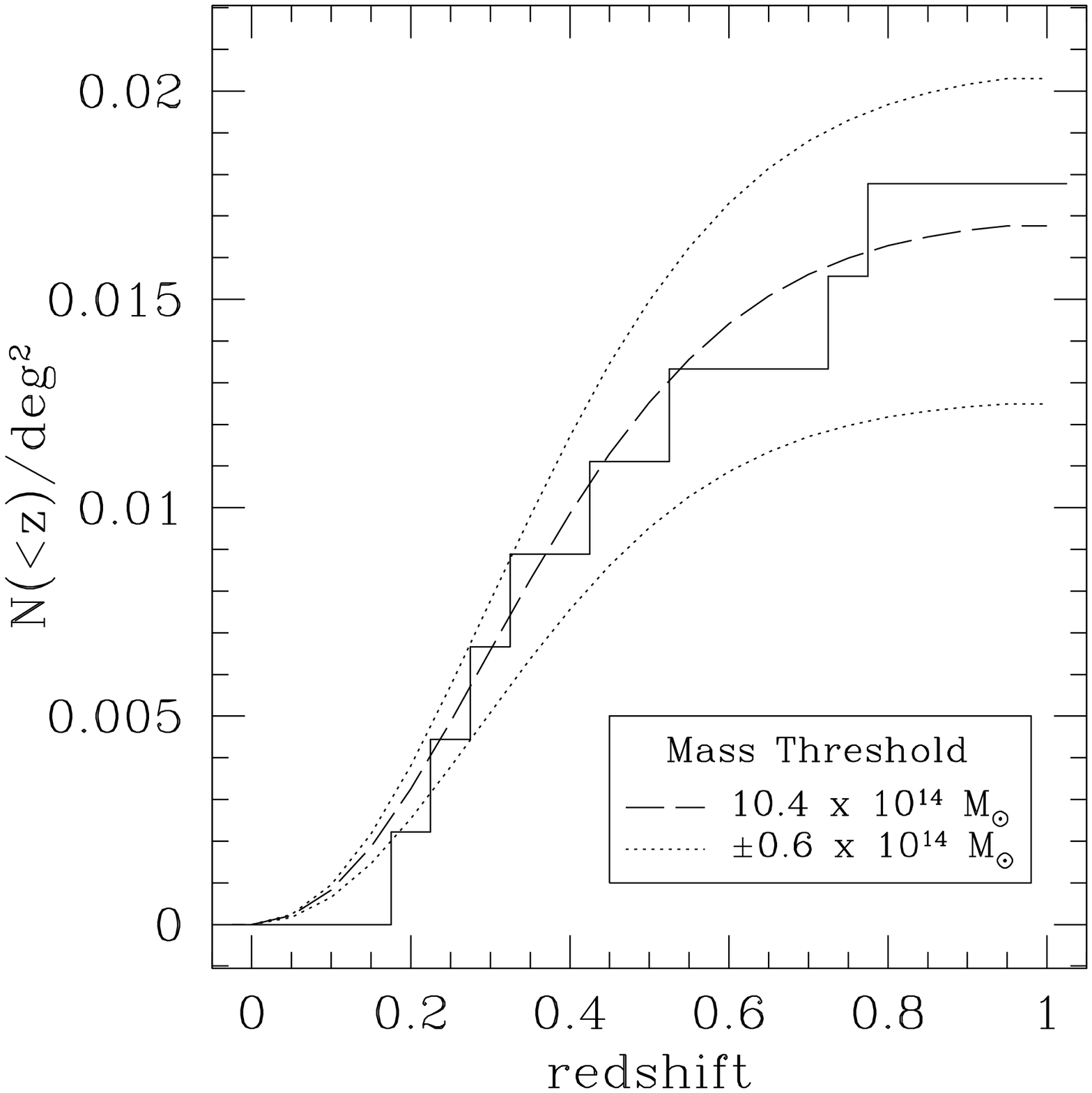}
\caption{ \label{fig:redshift_dist} The redshift distribution of the \totalClusters\ high-significance clusters listed in Table \ref{tab:clustProperties} compared to expectations from the mass function of \citet{Tinker2008} assuming all clusters above a given mass threshold have been detected.  Here we assume the total area observed is \area\ with uniform coverage, and the WMAP7+BAO+SN best-fit cosmology for a wCDM model \citep{Komatsu2010} for the mass function.  This figure suggests an effective mass threshold of our sample of $\approx 10.4 \times 10^{14} \Msun$ (dashed line).  For illustrative purposes we show via dotted lines the expected redshift distributions for mass thresholds larger/smaller by $0.6  \times 10^{14} \Msun$.} 
\end{figure}

We determine the selection function of our cluster subsample through both optical observations and simulations.  To investigate the effective mass threshold of our cluster sample, we plot cumulative clusters as a function of redshift and compare that to expectations from the mass function of \citet{Tinker2008} assuming the best-fit cosmology from WMAP7+BAO+SN with a wCDM model \citep{Komatsu2010}. 
We find that this sample is consistent with a mean mass\footnote{Note that throughout this text cluster masses are defined in terms of $M_{200}$, which is the mass within $R_{200}$, the radius within which the mean cluster density is 200 times the {\it{average}} density at the cluster redshift.} threshold of $10.4 \times 10^{14} \Msun$ for $M_{200}$ as shown by the dashed line in Figure \ref{fig:redshift_dist}.  For illustrative purposes, we show by dotted lines the expected redshift distributions for mass thresholds larger and smaller by $0.6 \times 10^{14} \Msun$, which enclose our subsample distribution for $z>0.25$.  This figure also assumes uniform coverage over an area of \area. Note that Figure \ref{fig:redshift_dist} is presented to give a qualitative understanding of our sample, but is not directly used in the cosmological analysis.

\begin{figure}[]
\epsscale{1.0}
\plotone{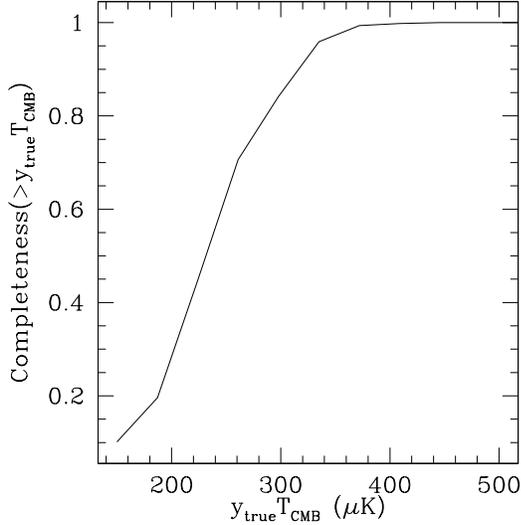}
\caption{\label{fig:completeness} Completeness versus true $yT_{\rm{CMB}}$ from simulations.  For an observed $yT_{\rm{CMB}}$ threshold of \ymin\ we expect our sample to be about $85\%$ complete for true $yT_{\rm{CMB}}$ values above \ymin.}
\end{figure}

\begin{figure}[t]
\epsscale{1.0}
\plotone{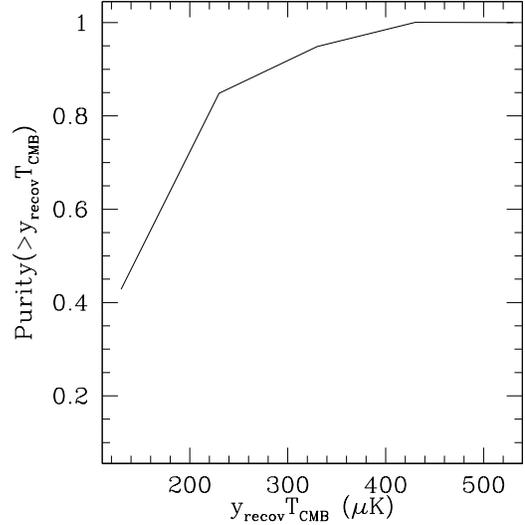}
\caption{\label{fig:falseDetections} Purity versus recovered $yT_{\rm{CMB}}$ from simulations.  For an observed $yT_{\rm{CMB}}$ threshold of \ymin\ we expect our sample to be about $90\%$ pure.  Our actual sample is $100\%$ pure, as each candidate cluster has been confirmed by optical observations \citep{Menanteau2010b}.}
\end{figure}

We also use simulations to characterize our sample.  The same cluster detection procedure discussed in \S \ref{subsec:detectMeth} is applied to the simulated millimeter-wave maps discussed in \S \ref{subsec:Yrecov}, including instrumental and atmospheric noise sources.  We find for an observed $yT_{\rm{CMB}}$ threshold of \ymin, we expect from simulations to be about 85\% complete above a true $yT_{\rm{CMB}}$ value greater than \ymin\  as shown in Figure  \ref{fig:completeness}.  This completeness is calculated as the total number of observed clusters with a recovered $y$ value above the threshold versus the total number of expected clusters with a true $y$ value above the threshold. This takes into account the scatter of observed $y$ values across the threshold.  We take \ymin\ as our threshold $yT_{\rm{CMB}}$ value and only consider clusters with measured $yT_{\rm{CMB}}$ values larger than this. 
We expect about 90\% purity for cluster detections with an observed $yT_{\rm{CMB}}$ value greater than \ymin.  Figure \ref{fig:falseDetections} illustrates how this false detection rate is expected to vary as a function of observed $yT_{\rm{CMB}}$ threshold.  From the optical observations discussed above, we find that all clusters identified in the millimeter-wave maps with $yT_{\rm{CMB}}$ values greater than \ymin\ were verified as clusters in the optical. Thus this sample is 100\% pure with no false detections (see \citet{Menanteau2010b}). 
Note that Figures \ref{fig:completeness}  and \ref{fig:falseDetections} are representative, coming from a relatively small sample of simulated high-significance clusters as discussed in \S \ref{subsec:Yrecov}.  They are presented to give a qualitative understanding of the sample and to understand above what $yT_{\rm{CMB}}$ threshold our sample is roughly complete.  Beyond this, they do not enter in the analysis of cosmological parameters.

\subsection{Scaling Relation Between SZ Signal and Mass}\label{sec:scaling}

We assume for the relation between the SZ signal and mass the general parameterized form\footnote{Note that the form of this relation is analogous to that in \citet{Vanderlinde2010}.}

\begin{equation}
\sntrue = A \bigg(\frac{\m}{M_0}\bigg)^B \bigg(\frac{1+z}{1+z_0}\bigg)^C,
\label{eqn:scaling}
\end{equation}
where $M_0 = 5 \times 10^{14} \Msunh$ and $z_0 = 0.5$.  We also assume that cluster $y$ values are randomly distributed around this relation with a lognormal scatter $\s$.  By lognormal scatter, we mean that the scatter in the relation ${\rm{ln}}(\sntrue) = {\rm{ln}}(A) + B{\rm{ln}}(\m/M_0) + C{\rm{ln}}((1+z)/(1+z_0))$ has a Gaussian distribution with a mean of zero and a standard deviation equal to $\s$.  This scaling relation relates the observable quantity for each cluster, SZ signal, to the quantity most directly relevant for cosmology, the cluster mass. This form is chosen as the SZ signal is expected to have a power-law dependence on mass.  Some dependence of our recovered SZ signal on redshift is allowed with an additional parameter $C$.  Note that the $B$ and $C$ parameters are independent of $M_0$ and $z_0$, whose values only affect $A$.  We also note that the uncertainty in the relation between SZ signal and mass is dominated by astrophysical processes and is only minimally dependent on cosmological parameters given WMAP7 priors.

\begin{figure}[t]
\epsscale{1.1}
\plotone{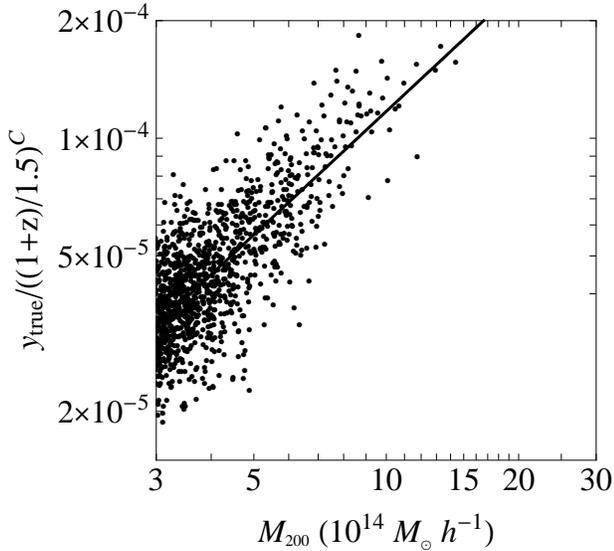}
\caption{\label{fig:scalingRelation} Relation between true $y$ and the cluster mass from simulations, including clusters with $M_{200} > 3 \times 10^{14} \Msunh$ and $z > 0.15$.  The best-fit scaling relation parameters in Eq.~\ref{eqn:scaling} are found with a least-squares fit, and the resulting best-fit surface is plotted in two-dimensions as the solid line above.  The lognormal scatter between the true $y$ values and the best-fit scaling relation is $26\%$.}
\end{figure}

We use the simulations discussed in \S \ref{subsec:Yrecov} to determine fiducial model values for $A$, $B$, and $C$.  In particular, we use only the simulated maps containing the first-order thermal SZ component of the simulated clusters, without altering the maps by adding any noise or convolving with the ACT beam.  Using the $y$ value of the brightest $0.5 \arcmin$ pixel for each cluster in the simulated thermal SZ maps, we solve for the best-fit values of the three scaling relation parameters in Eq.~\ref{eqn:scaling}, as well as the scatter in this relation, with a linear least squares fit.  Figure \ref{fig:scalingRelation} shows the simulated clusters along with the best-fit line, and Table \ref{tab:scalingRelation} gives the corresponding best-fit values and errors of the fiducial scaling relation parameters in addition to the scatter $S$.  The best-fit values for the $B$ and $C$ parameters are close to what we would expect from self-similar scaling relations as discussed in Appendix \ref{app:scaling}.

There are a number of astrophysical mechanisms that could cause the observed relation between SZ signal and mass to differ from the fiducial relation.  One is contamination of SZ decrements by radio or infrared galaxies at 148 GHz.  Regarding radio galaxies, observations suggest that these galaxies show some preference for residing in galaxy clusters \citep{Coble2007,Lin2007,Lin2009,Mandelbaum2009}.  However, using a model of radio galaxies that describes their correlation with halos, the amount of contamination expected from radio galaxies was found to be negligible for  \citep{Sehgal2010}. For redshifts $< 1$, star formation, which is responsible for infrared galaxy emission, is expected to be quenched in high-density environments.  At low redshifts ($z\sim0.06$) the fraction of all galaxies that are star forming galaxies is $\sim 16\%$ in clusters \citep{Bai2010}.  While this percentage is expected to increase at higher redshifts, given that the total infrared background at 150 GHz is roughly 30 $\rm{\mu K}$ \citep{Fixen1998}, it is unlikely that infrared galaxy contamination could be significant for clusters with $yT_{\rm{CMB}} > 300~\mk$.  \citet{Lima2010} have also shown that the lensing of infrared galaxies by massive clusters should not introduce a significant bias in the measured SZ signals.

Another way for the observed SZ signal to be lower than the fiducial model is if clusters have a significant amount of nonthermal pressure.  This pressure would not be observed as part of the SZ signal, however, it would play an important role in counteracting the gravitational pressure from the cluster mass.  Such nonthermal pressure can take the form of small scale turbulence, bulk flows, or cosmic rays.  Simulations and observations suggest contributions to the total pressure from cosmic rays to be about $5-10\%$ \citep{Jubelgas2008, Pfrommer2004} and from turbulent pressure to be between $5-20\%$ \citep{Lau2009, Meneghetti2010, Burns2010},  with only the latter work suggesting levels as high as $20\%$ and that largely at the cluster outskirts.  These processes have a much larger impact on lower mass clusters and groups where the gravitational potential is not strong enough to tightly bind the cluster gas \citep[e.g.,][]{Battaglia2010,Shaw2010,Trac2010}. However, for the massive systems considered here, this again is not expected to be a significant issue.  One astrophysical process that can have a significant affect on the cluster $y$ values  is major mergers.  We certainly have at least one in our sample (Bullet cluster), but note the extreme rarity of such objects in general.

\begin{figure*}[ht]
\begin{center}
\includegraphics[scale=0.56]{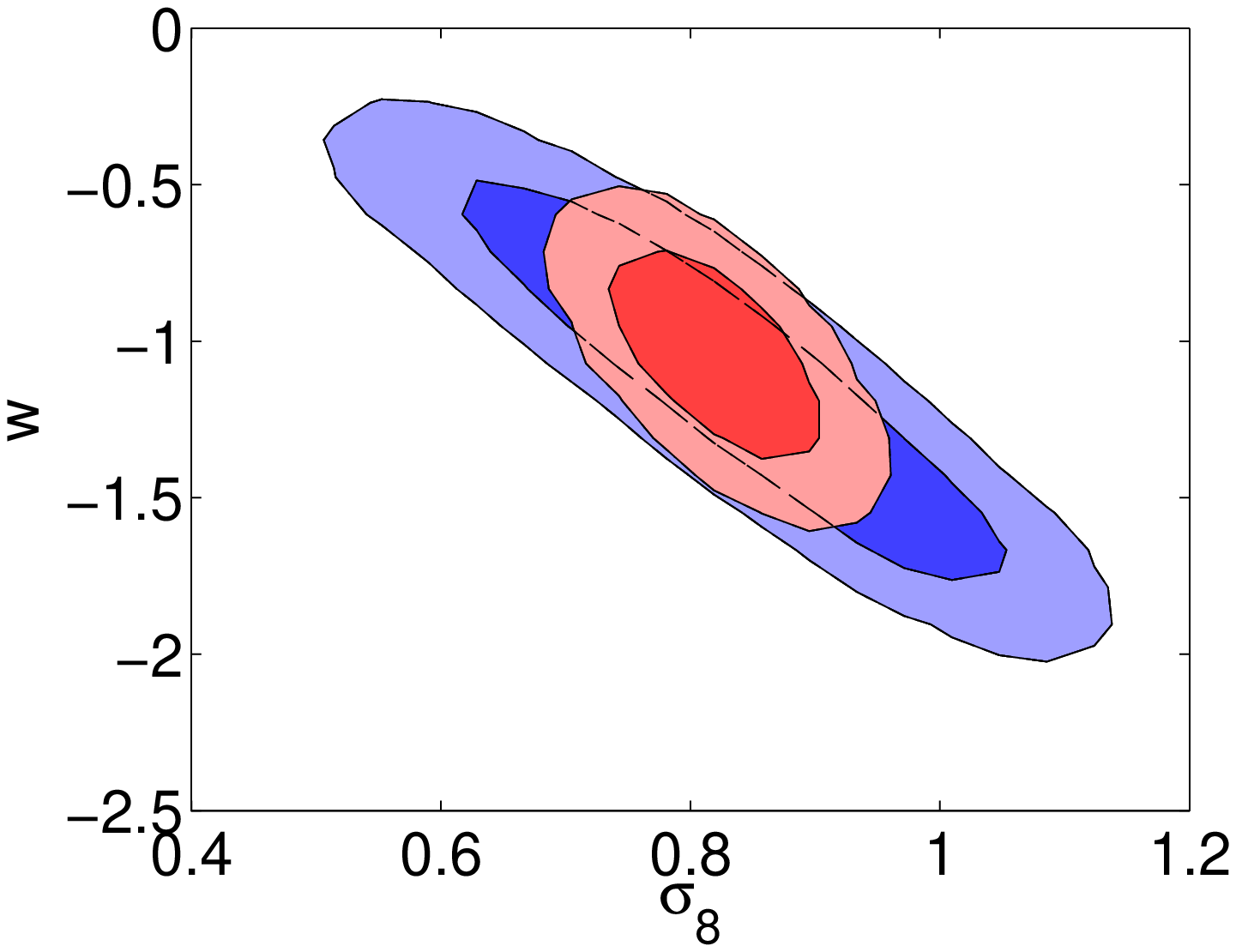}\includegraphics[scale=0.56]{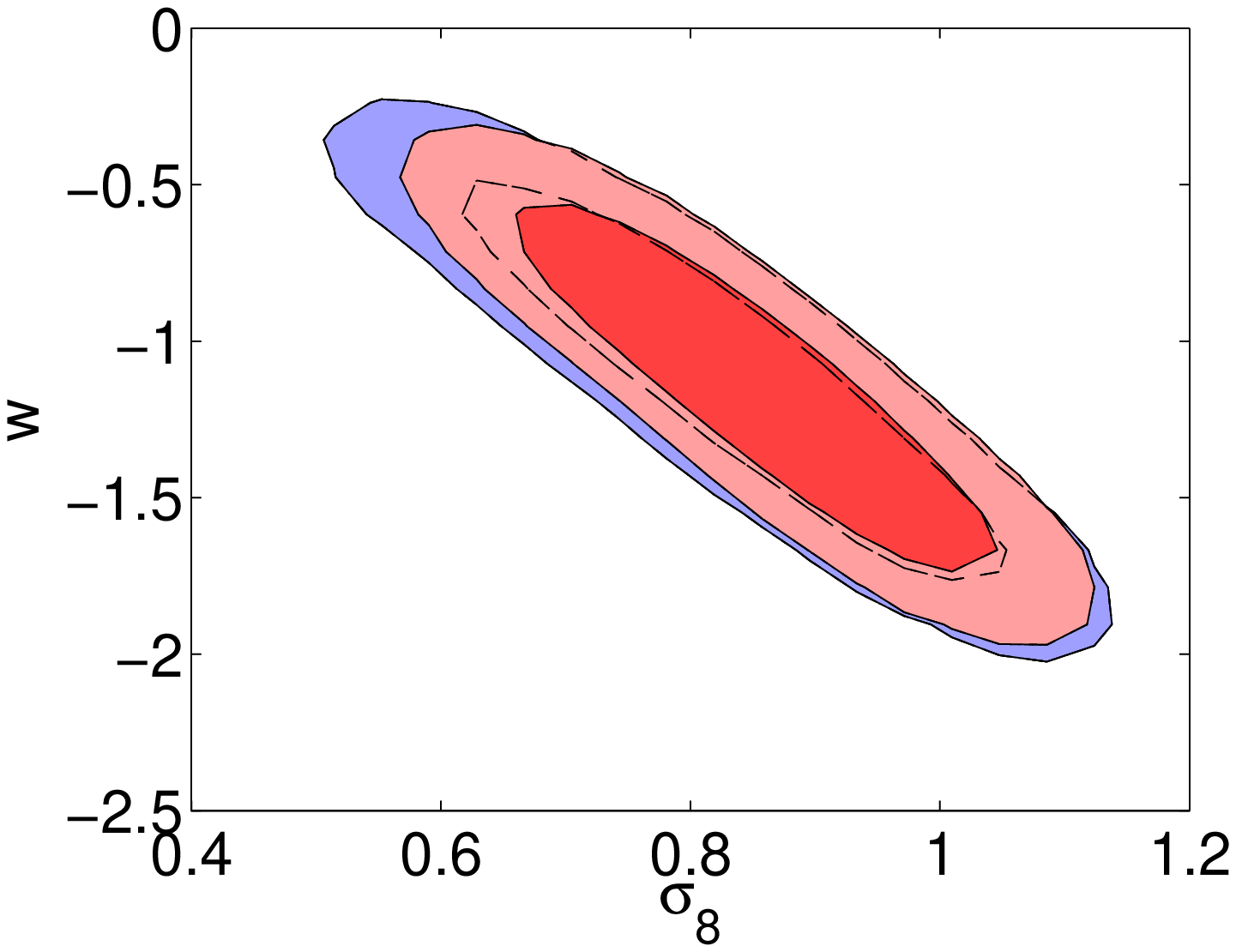}
\caption{Likelihood contour plots of $w$ versus $\sigma_8$ showing 1$\sigma$ and 2$\sigma$ marginalized contours.  {\emph{Left}:} Blue contours are for WMAP7 alone, and red contours are for WMAP7 plus ACT SZ detected clusters, fixing the mass-observable relation to the fiducial relation given in \S  \ref{sec:scaling}.  {\emph{Right:}} Contours are the same as in the left panel, except that the uncertainty in the mass-observable relation has been marginalized over within priors discussed in \S \ref{sec:likelihood}.}
\label{fig:constraints}
\end{center}
\end{figure*}

\begin{center}
\begin{deluxetable*}{ccccc}
\tabletypesize{\scriptsize}
\tablecaption{Best-fit Scaling Relation Parameters\label{tab:scalingRelation}}
\tablehead{\colhead{Model and Data Set} & \colhead{A} & \colhead{B} & \colhead{C} & \colhead{S}}
\startdata
Simulation Fiducial Values & $(5.67 \pm 0.05)\times 10^{-5}$ & $1.05 \pm 0.03$ & $1.29 \pm 0.05$ & 0.26\\
wCDM WMAP7 + ACT Clusters & $(\Aparam)\times 10^{-5}$ & $\Bparam$ & $\Cparam$ & $\Sparam $
\enddata
\end{deluxetable*}
\end{center}

\begin{center}
\begin{deluxetable*}{ccc}
\tabletypesize{\scriptsize}
\tablecaption{Cosmological Parameter Constraints for $\sigma_8$ and $w$
\label{tab:constraints}}
\tablehead{\colhead{Model and Data Set} & \colhead{$\sigma_8$} & \colhead{w} }
\startdata
wCDM WMAP7+BAO+SN & $0.802\pm0.038$ &  $-0.98 \pm 0.053$\\
wCDM WMAP7 & $0.835 \pm 0.139$ & $-1.11 \pm 0.40$\\
wCDM WMAP7 + ACT Clusters (fiducial scaling relation)& $ \signomarg$ & $\wnomarg$\\
wCDM WMAP7 + ACT Clusters (marginalized over scaling relation) & $ \sigwmarg$ & $\wwmarg$
\enddata
\end{deluxetable*}
\end{center}

\vspace{-1.2cm}

\subsection{Cluster Likelihood Function}\label{sec:likelihood}

In order to constrain cosmological parameters with our cluster sample, we construct a likelihood function specific for clusters, and we map out the posterior distribution to find marginalized distributions for each parameter.  We follow \citet{Cash1979} who derived the likelihood function in the case of Poisson statistics giving

\begin{equation}
{\rm{ln}}\mathcal{L} = {\rm ln}\pr(\{ n_i \}| \{ \lambda_i \}) = \sum_{i=1}^{N_b}  ({n_i \rm{ln}} \lambda_i -  \lambda_i).
\end{equation}
$\pr$ is the probability of measuring $\{ n_i \}$ given modeled counts $\{\lambda_i \}$.  Here $N_b$ is the total number of observed bins in SZ signal - redshift space, and $\lambda_i$ is the modeled number of clusters in the $i$th bin.  We also take the bin sizes to be small enough so that no more than one observed cluster is in each bin.  The modeled cluster count,  $\lambda_i$, is a function of the SZ signal and redshift of the given bin (which we call $\snobs$ and $\zobs$) as well as the set of cosmological parameters, $\cosm$.  The modeled count is also a function of the parameters of the SZ signal - mass scaling relation ($A, B, C, S$) given in Eq.~\ref{eqn:scaling}, since it is the abundance of clusters as a function of mass that is tied to cosmology via the mass function.  For this work we use the mass function given in \citet{Tinker2008}.  A derivation of the full cluster likelihood function used in this analysis can be found in Appendix \ref{app:like}.  This likelihood is given by Eq.~\ref{eqn:fullLike} and is a function of the parameters $\cosm$ and $A, B, C, S$.  

We assume normal errors of $2.2\times 10^{-5}$ on $\snobs$ (corresponding to an error in $yT_{\rm CMB}$ of 60 $\mu$K) and 0.1 on $\zobs$. We take 0.1 as the redshift uncertainty for convenience even though six of our clusters have spectroscopic redshifts.  However, the redshift error does not dominate the uncertainty of our results.  We also assume Gaussian priors on $A, B, C,$ and $S$ centered around the fiducial values given in Table \ref{tab:scalingRelation}, with conservative $1\sigma$ uncertainties of $35\%, 20\%, 50\%,$ and $20\%$ respectively of the fiducial values.  These priors were determined by finding the relation between SZ signal and mass from simulated thermal SZ maps with varying gas models.  In particular, we use two simulated thermal SZ maps analogous to those discussed in \S \ref{sec:scaling}, with the gas physics models in these maps based on the adiabatic and the nonthermal20 models described in \citet{Trac2010}.  The adiabatic model assumes no feedback, star-formation, or other nonthermal processes that could lower the SZ signal as a function of mass.  The nonthermal20 model assumes more star-formation than the fiducial model and $20\%$ nonthermal pressure support for all clusters at all radii, which is a larger amount of nonthermal pressure than generally suggested by X-ray observations and hydrodynamic simulations \citep[e.g.,][]{Lau2009, Meneghetti2010,Burns2010}.  These two models span the range of plausible gas models for massive clusters given current observations, and the $1\sigma$ priors on the scaling relation parameters given above are generous given the range in parameters spanned by these models.  

\subsection{Parameter Constraints}\label{sec:params}

The likelihood function described above was made into a standalone code module which was then interfaced with the Markov chain software package CosmoMC \citep{Lewis2002}.  Using CosmoMC, we run full chains for the WMAP7 data alone \citep{Larson2010} and for the WMAP7 data plus our ACT cluster subsample.  We assume a wCDM cosmological model which allows $w$ to be a constant not equal to $-1$, assumes spatial flatness, and which has as free parameters: $\Omega_b h^2$,  $\Omega_c h^2$, $\theta_*$, $\tau$, $w$, $n_s$, ${\rm{ln}}[10^{10} A_s]$, and $A_{\rm{SZ}}$ as defined in \citet{Larson2010}.  The parameter $\sigma_8$ is derived from the first seven of these parameters and is kept untied to $A_{\rm{SZ}}$ as the link between the two is in part what we are investigating.  We run the WMAP7 plus ACT clusters chain under two cases: one where the values of $A, B, C,$ and $S$ are fixed to the fiducial values given in \S \ref{sec:scaling} and listed in Table \ref{tab:scalingRelation}, and one where $A, B, C,$ and $S$ are allowed to vary within the conservative priors given in \S  \ref{sec:likelihood}.  For the latter case, we add these four new parameters to the CosmoMC code.  At each step of the chain, CosmoMC calls the software package CAMB\footnote{www.camb.info} to generate both the microwave background power spectrum and matter power spectrum as a function of the input cosmology, and then the natural logarithms of both the WMAP and cluster likelihoods are added.  We determine the posterior probability density function through the Markov chain process and use a simple R$-$1 statistic \citep{Gelman1992} of R$-$1 < 0.01 to check for convergence of the chains.   

The best-fit marginalized 1$\sigma$ and 2$\sigma$ contours, obtained from this process, are shown in Figure \ref{fig:constraints} for $w$ and $\sigma_8$.  The blue contours show the constraints for WMAP7 alone, while the red contours show the constraints from the union of WMAP7 plus our ACT cluster subsample.  The left panel shows the best-fit contours with the SZ signal - mass scaling relation fixed to the fiducial relation obtained from the simulations.  The right panels show the constraints allowing the four parameters of the scaling relation to vary.  Table \ref{tab:constraints} lists the best-fit parameter values for $\sigma_8$ and $w$ with their $1\sigma$ marginalized uncertainties.  Table \ref{tab:scalingRelation} lists the best-fit scaling relation values and $1\sigma$ uncertainties as well as the fiducial values obtained from simulations as discussed in \S \ref{sec:scaling} for comparison.  We note that for the remaining seven parameters fit in the analyses combining WMAP7 plus ACT clusters ($\Omega_b h^2$,  $\Omega_c h^2$, $\theta_*$, $\tau$, $n_s$, ${\rm{ln}}[10^{10} A_s]$, and $A_{\rm{SZ}}$), we find best-fit values consistent with the best-fit values from WMAP7 alone with a modest improvement in the marginalized errors.

%\vspace{0.07cm}
\subsection{Stacked SZ Signal}\label{sec:stacking}

\begin{figure}[t]
\epsscale{1.1}
\plotone{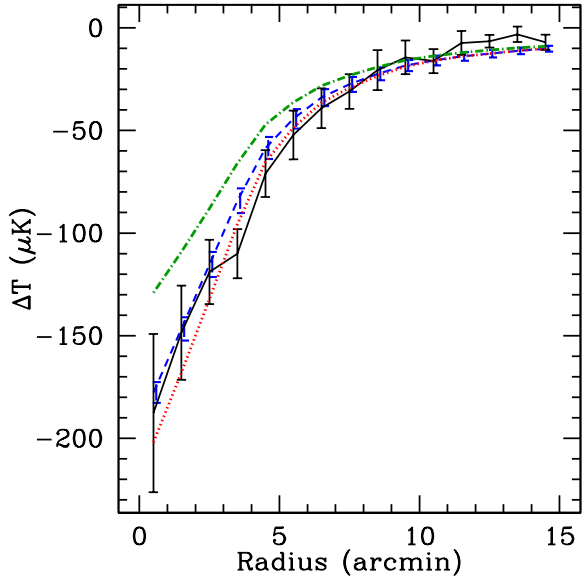}
\caption{\label{fig:stackedClusters} Average profile from stacking the 9 clusters presented in Table \ref{tab:clustProperties} (black solid line) compared with the average profile stacking 40 clusters with unsmoothed $yT_{\rm{CMB}}$ values greater than \ymin\ in simulated thermal SZ maps convolved with the ACT beam (dashed blue line).  The dashed blue line shows the average profile for clusters simulated with the fiducial SZ model, while the dotted red (bottom) and dot-dashed green (top) lines show the same assuming the adiabatic and nonthermal20 SZ models respectively which are discussed in \S \ref{sec:likelihood}.  For the profiles of the 9 clusters in the data, we removed a mean background level from the profile of each cluster.  Error bars for the simulated clusters have been offset by $0.1\arcmin$ for clarity, and are smaller than those from the data.  Error bars for the adiabatic and nonthermal20 models are not shown, but are of similar size as for the fiducial model.}
\end{figure}

We also perform a stacking analysis of the nine clusters listed in Table 1 to
measure average cluster SZ profiles, which can be compared with simulations.
We stack the 9 clusters in the data map prior to any filtering, after subtracting a mean background level for each cluster profile using an annulus 15$^\prime$ from the center of each cluster and $0.5 \arcmin$ wide.  The stacked average profile is given by the solid black line in Figure \ref{fig:stackedClusters}.  The same procedure is preformed on all simulated clusters with unsmoothed $yT_{\rm{CMB}}$ values greater than \ymin\ in simulated thermal SZ maps.  There are 40 of these simulated clusters in total over 6 different 455 square degree maps spanning the same redshift range as the data.  These simulated clusters are stacked in thermal SZ maps convolved with the ACT beam to mimic the data, and their average profile is given by the dashed blue line in Figure \ref{fig:stackedClusters}.  Error bars represent the standard deviation of the mean in each radial bin.  The blue dashed line represents the stacked profiles of simulated clusters assuming the fiducial SZ model.  The red dotted and green dot-dashed lines show the stacked profiles of simulated clusters assuming the adiabatic and nonthermal20 SZ models discussed in \S \ref{sec:likelihood}.  The error bars have not been included for the latter two models in Figure \ref{fig:stackedClusters}, but they are of similar size as for the fiducial model. We find good agreement in the average profiles of the clusters in the data and simulated with the fiducial model as shown in Figure \ref{fig:stackedClusters}, which suggests that there is no significant misestimate of the SZ signal for these massive systems. 

\section{DISCUSSION}\label{sec:discuss}

From Table \ref{tab:constraints} we see overall agreement between $\sigma_8$ and $w$ as measured with only WMAP7 and as measured with the high-significance ACT cluster sample plus WMAP7.  We find $\sigma_8 = \signomarg$ and $w=\wnomarg$ if we assume the fiducial scaling relation, a decrease in the uncertainties on these parameters by roughly a factor of three and two respectively as compared to WMAP7 alone.  This indicates the potential statistical power associated with cluster measurements.  Marginalizing over the uncertainty in this scaling relation, we find $\sigma_8 = \sigwmarg$ and $w=\wwmarg$, an uncertainty comparable to that of WMAP7 alone.  We also see consistency when comparing these constraints to the best-fit constraints from WMAP7 plus baryon acoustic oscillations plus type Ia supernovae, which give $\sigma_8 = 0.802 \pm 0.038$ and $w=-0.980 \pm 0.053$ for a wCDM cosmological model \citep{Komatsu2010}.  As the latter are all expansion rate probes, this suggests agreement between expansion rate and growth of structure measures.  Both also show $w$ is consistent with $-1$, giving further support to dark energy being an energy of the vacuum.  

These results are also consistent with analyses from X-ray cluster samples giving $\sigma_8(\Omega_m/0.25)^{0.47} = 0.813 \pm 0.013$ (stat) $\pm 0.024$ (sys) and $w=-1.14 \pm 0.21$ \citep{Vikhlinin2009b}, $\Omega_m = 0.23 \pm 0.04$, $\sigma_8 = 0.82 \pm 0.05$, and $w= -1.01\pm 0.20$ for a wCDM model \citep{Mantz2010}, and $\Omega_m = 0.30^{+0.03}_{-0.02}$, $\sigma_8 = 0.85^{+0.04}_{-0.02}$ from WMAP5 plus X-ray clusters \citep{Henry2009} .  We also find consistency with optical samples yielding $\sigma_8(\Omega_m/0.25)^{0.41} = 0.832 \pm 0.033$ for a flat $\Lambda$CDM model \citep{Rozo2010}.  \citet{Vanderlinde2010} find $\sigma_8 = 0.804 \pm 0.092$ and $w=-1.049 \pm 0.291$ for a wCDM model using SZ clusters detected by SPT plus WMAP7. 

This analysis also suggests consistency between the fiducial model of cluster astrophysics used here to describe massive clusters and the data.  Table \ref{tab:constraints} shows agreement in best-fit cosmological parameters between growth rate and expansion rate probes when we hold fixed our fiducial relation between SZ signal and mass.  When we allow the scaling relation parameters to be free, we find best-fit values that are broadly consistent with those of our fiducial relation.  We note that while the $1\sigma$ range of the $B$ parameter is higher than the fiducial value, the fiducial value is enclosed by the $2\sigma$ range of $1.75^{+0.4}_{-0.7}$.  The higher value of the $B$ parameter may indicate some curvature in the true scaling relation away from the fiducial model at the high-mass end.  This may also be suggested by Figure \ref{fig:scalingRelation} where the simulated clusters seem to prefer higher $y$ values than the fiducial relation would suggest for the most massive systems.  The agreement between cosmological parameters from expansion rate and growth of structure probes when fixing the SZ signal - mass scaling relation to the fiducial model and the broad agreement between fiducial and best-fit scaling relation parameters when the latter are allowed to be free, suggest our data is broadly consistent with expectations for  the SZ signal of massive clusters.  This is also suggested by comparing the stacked SZ detected clusters in the data with simulations as shown in Figure \ref{fig:stackedClusters}.  

We would expect the above to be the case as massive clusters have been studied far better than lower mass clusters with a variety of multi-wavelength observations.  In addition, a number of astrophysical processes that are not perfectly understood, such as nonthermal processes and point source contamination, affect the gas physics of lower mass clusters much more than that of the most massive systems.  In general, these processes tend to suppress the SZ power spectrum over that of a straightforward extrapolation based on the most massive systems. This is an important effect as lower mass systems ($< 10^{14} \Msun$) contribute as much to the SZ power spectrum at $l\sim3000$ as systems at higher mass \citep{Komatsu2002,Trac2010}.
The power spectrum near $l=3000$ has been recently measured by and discussed in \citet{Lueker2009}, \citet{Das2010} and \citet{Dunkley2010}.

There are a number of ways the cosmological constraints presented here could be further improved.  Clearly the largest uncertainty is the relation between SZ signal and mass, and further X-ray observations of massive clusters, particularly at higher redshifts where X-ray observations have been limited, would help to calibrate this relation.  Further targeted observations of massive clusters at millimeter-wave frequencies with enough resolution and sensitivity to identify point  sources would offer a better handle on contamination levels.  In addition, an analysis using multiple frequency bands, which would employ the spectral information of the SZ signal, may be helpful in determining cluster sizes and measuring integrated $Y$s.  This could help reduce the scatter in the relation between SZ signal and mass.  Spectroscopic redshifts of all the clusters in a given SZ sample would also help to reduce uncertainty on the cosmological parameters.   In addition, millimeter-wave maps with lower instrument noise, would greatly reduce the scatter between the recovered and true SZ signal.  Such maps are expected with ACTpol \citep{Niemack2010} and SPTpol \citep{McMahon2009} coming online in the near future.  

With continued SZ surveys such as ACT and SPT and their polarization counterparts, in addition to data forthcoming from the {\it{Planck}} satellite, we will no doubt increase the number of SZ cluster detections.  We anticipate that upcoming larger galaxy cluster catalogs will make significant contributions to our understanding of both cluster astrophysics and cosmology.

\acknowledgments
NS would like to thank Phil Marshall for very insightful discussions regarding the construction of the likelihood function and Adam Mantz and David Rapetti for many helpful discussions, particularly in regard to CosmoMC.  NS also acknowledges useful conversations with Michael Busha, Glenn Morris, Jeremy Tinker, and Roberto Trotta.  

This work was supported by the U.S. National Science Foundation through awards AST-0408698 for the ACT project, and PHY-0355328, AST-0707731 and PIRE-0507768. Funding was also provided by Princeton University and the University of Pennsylvania. The PIRE program made possible exchanges between Chile, South Africa, Spain and the U.S. that enabled this research program. Computations were performed on the GPC supercomputer at the SciNet HPC Consortium. SciNet is funded by: the Canada Foundation for Innovation under the auspices of Compute Canada; the Government of Ontario; Ontario Research Fund -- Research Excellence; and the University of Toronto.

NS is supported by the U.S. Department of Energy contract to SLAC no. DE-AC3-76SF00515.  AH, TM, SD, and VA were supported through NASA grant NNX08AH30G. AH received additional support from a Natural Science and Engineering Research Council of Canada (NSERC) PGS-D scholarship.  AK and BP were partially supported through NSFAST-0546035 and AST-0606975, respectively, for work on ACT.  ES acknowledges support  by NSF Physics Frontier Center grant PHY-0114422 to the Kavli Institute of Cosmological Physics. HQ and LI acknowledge partial support from FONDAP Centro de Astrofisica. JD received support from an RCUK Fellowship. KM, MH, and RW received financial support from the South African National Research Foundation (NRF), the Meraka Institute via funding for the South African Centre for High Performance Computing (CHPC), and the South African Square Kilometer Array (SKA) Project. RD was supported by CONICYT, MECESUP, and Fundacion Andes. RH acknowledges funding from the Rhodes Trust. SD acknowledges support from the Berkeley Center for Cosmological Physics. YTL acknowledges support from the World Premier International Research Center Initiative, MEXT, Japan.   Some of the results in this paper have been derived using the HEALPix package \citep{Gorski2005}.  We acknowledge the use of the Legacy Archive for Microwave Background Data Analysis (LAMBDA). Support for LAMBDA is provided by the NASA Office of Space Science. The data will be made public through LAMBDA (http://lambda.gsfc.nasa.gov/) and the ACT website (http://www.physics.princeton.edu/act/).

\appendix

\section{A. Self-Similar Scaling Relation Between SZ Signal and Mass}
\label{app:scaling}

If clusters were self-similar and isothermal, then we would expect the scaling relation between SZ signal and mass to be
\begin{equation}
Y_{\rm{halo}} \propto M_{\rm{halo}}^{5/3} E(z)^{2/3} f_{\rm{gas}}/d_A^2,
\end{equation}
where $Y_{\rm halo}$ is the Compton y-parameter integrated over the surface of the cluster in units of arcmin$^2$, and $E(z)=[\Omega_m (1 + z)^3 + \Omega_{\Lambda}]^{1/2}$ for a flat $\Lambda \rm{CDM}$ cosmology. The angular diameter distance is denoted by $d_A$, and $f_{\rm{gas}}$ is the gas mass fraction.  For the Compton y-parameter integrated over a fixed aperture we have
\begin{equation}
Y_{\rm{aperture}} \propto Y_{\rm{halo}} \left(\frac{R_{\rm aperture}}{R_{\rm halo}} \right)^2,
\end{equation}
where $R_{\rm aperture} \propto d_A$ and $R_{\rm halo}\propto M_{\rm halo}^{1/3} E(z)^{-2/3}$.  Note that this equation is appropriate if the aperture size is smaller than the size of the cluster. 
The above gives
\begin{equation}
Y_{\rm{aperture}} \propto M_{\rm{halo}} E(z)^2 f_{\rm{gas}}. 
\end{equation}
To write $Y_{\rm{aperture}}$ as a function of $(1+z)$ we note that at z=0.5 (the mean redshift of our cluster sample) $E(z) \propto (1+z)^{0.835}$ for $\Omega_m = 0.27$.  Thus 
\begin{equation}
Y_{\rm{aperture}} \propto M_{\rm halo}^B \, (1+z)^C,
\end{equation}
where $B = 1.0$ and $C=1.67$.

\section{B. Cluster Likelihood Function}
\label{app:like}

Below we describe the construction of the likelihood function for SZ clusters detected in millimeter-wave surveys.  From Poisson statistics, the probability of observing $n_i$ counts expecting $\lambda_i$ counts is

\begin{equation}
\pr(n_i | \lambda_i) = \frac{\lambda_i^{n_i} e^{-\lambda_i}}{n_i!}.
\end{equation}
Given a data set of $\{n_i\}$ counts in $N_b$ observed bins and a corresponding prediction, $\{\lambda_i\}$, the probability of the data given the prediction is

\begin{equation}
\pr(\{ n_i \}| \{ \lambda_i \}) =  \prod_{i=1}^{N_b} \frac{\lambda_i^{n_i} e^{-\lambda_i}}{n_i!}, 
\end{equation}
where $\lambda_i = \pr(\snobs,\zobs, A, B, C, \s, \cosm)N \Delta\snobs \Delta\zobs $.   Here $\pr(\snobs,\zobs, A, B, C, \s, \cosm)$ is the probability of observing a cluster in bin 
$i$, and $N$ is a normalization factor giving $\lambda_i$ units of counts (see below).  The observed SZ signal and redshift of a given cluster are denoted by $\snobs$ and $\zobs$, and 
$\Delta\snobs$ and $\Delta\zobs $ denote the size of the bin.  The parameters $A, B, C, \s$ describe the scaling relation between SZ signal and mass and are defined below.  The cosmological 
parameters are indicated by $\cosm$.  

If we allow the bin sizes to be small enough that each observed bin holds no more than one observed cluster, then

\begin{equation}
{\rm ln}\pr(\{ n_i \}| \{ \lambda_i \}) = \sum_{i=1}^{n}  {\rm{ln}} \lambda_i - \sum_{i=1}^{N_b} \lambda_i = {\rm{ln}}\mathcal{L}
\end{equation}
as given in \citet{Cash1979}. Note that the ${\rm{ln}}(n_i!)$ term has been dropped as it is independent of any change in parameters, and $n$ represents the total number of clusters observed. 
Thus we have

\begin{equation}
{\rm{ln}}\mathcal{L} =  \sum_{i=1}^{n} {\rm ln}(\pr(\snobs,\zobs, A, B, C, \s, \cosm) N d\snobs d\zobs) - \int_{\zobs} \int_{\snobs} \pr(\snobs,\zobs, A, B, C, \s, \cosm) N d\snobs d\zobs 
\end{equation}
with

\begin{equation}
\pr(\snobs,\zobs, A, B, C, \s, \cosm) = \iiint \pr(\snobs,\zobs, A, B, C, \s, \cosm, \sntrue, \ztrue, \lnm) d\sntrue d\ztrue d\lnm
\end{equation}

\begin{equation}
=  \iiint \pr(\snobs,\zobs | A, B, C, \s, \cosm, \sntrue, \ztrue, \lnm) \pr(A, B, C, \s, \cosm, \sntrue, \ztrue, \lnm) d\sntrue d\ztrue d\lnm
\end{equation}

\begin{equation}
=  \iiint  \pr(\snobs,\zobs | \sntrue, \ztrue) \pr(\sntrue | A, B, C, \s, \cosm,  \ztrue, \lnm) \pr(A, B, C, \s, \cosm,  \ztrue, \lnm) d\sntrue d\ztrue d\lnm
\end{equation}

\begin{eqnarray}
& & =  \int_{-\infty}^{\infty}  d\lnm \int_{0}^{\infty} d\sntrue  \int_{0}^{\infty} d\ztrue \pr(\snobs | \sntrue) \pr(\zobs | \ztrue) \pr(\sntrue | A, B, C, \s,  \ztrue, \lnm) \pr(\lnm, \ztrue | \cosm)  \nonumber \\ 
& &\hspace{8cm} \times \; \pr(\cosm) \pr(A) \pr(B) \pr(C) \pr(\s) 
\label{eqn:fullLike}
\end{eqnarray}
using the definition of conditional probability.  Here $\pr(\cosm)$ is any external prior on $\cosm$ such as a WMAP prior.

We assume the following SZ signal - mass scaling relation with log normal scatter, $\s$,

\begin{equation}
\sntrue = A \bigg(\frac{\m}{M_0}\bigg)^B \bigg(\frac{1+z}{1+z_0}\bigg)^C,
\end{equation}
where $M_0 = 5 \times 10^{14} \Msunh$ and $z_0 = 0.5$.  This gives

\begin{equation}
\pr(\sntrue | A, B, C, \s, \ztrue, \lnm) = \frac{1}{\sqrt{2\pi} \s \sntrue} \exp \bigg( \frac{- ({\rm ln} \sntrue - B \lnm - C {\rm ln} (1+\ztrue) - {\rm ln} A + B  {\rm ln} M_0 + C {\rm ln} (1+z_0))^2}{2 \s^2} \bigg) .
\end{equation}
We also assume Gaussian priors on the scaling relation parameters as indicated by simulations, giving

\begin{equation}
\pr(A) = \frac{1}{\sqrt{2\pi} \sigma_{\rm{A}}} \exp \bigg( \frac{- (A - A_0)^2}{2\sigma_{\rm{A}}^2} \bigg).
\end{equation}
Similar relations hold for $\pr(B)$, $\pr(C)$, and $\pr(\s)$.

From the mass function we have

\begin{equation}
\pr(\lnm, \ztrue | \cosm) = \frac{dn (\lnm, \ztrue, \cosm)}{d\lnm} \frac{dV(\ztrue,\cosm)}{d\ztrue} \frac{1}{N},
\end{equation}
where here $n$ is the number density of clusters.  $N$ is the total number of clusters when the above mass function 
is integrated over $d\lnm$ and $d\ztrue$. 

We also assume for the uncertainty on the observed SZ signal and redshift that

\begin{equation}
\pr(\snobs | \sntrue) = \frac{1}{\sqrt{2\pi} \sigma_{\rm{obs}}} \exp \bigg( \frac{- (\snobs - \sntrue)^2}{2\sigma_{\rm{obs}}^2} \bigg) 
\label{eq:sn}
\end{equation}

\begin{equation}
\pr(\zobs | \ztrue) = \frac{1}{\sqrt{2\pi} \sigma_{\rm{z}}} \exp \bigg( \frac{- (\zobs - \ztrue)^2}{2\sigma_{\rm{z}}^2} \bigg)
\end{equation}
where these two expressions should also be multiplied by $\frac{2}{1+{\rm erf}(x^{\rm true}/\sqrt{2}\sigma_x)}$ since the limits of integration are from 0 to $\infty$.  \\

\vspace{-0.2cm}

\bibliographystyle{hapj}

%\bibliography{refs}

\end{document}